\documentclass[aps,prd,onecolumn,groupedaddress,showpacs,nofootinbib,amssymb]{revtex4}
\usepackage[dvips]{graphicx}
\usepackage{amssymb}
\usepackage{amsmath}
\usepackage{graphicx,,color}
\usepackage{amsfonts}
\usepackage{bm}
\usepackage{cancel}
\usepackage{comment}

\newcommand\be{\begin{equation}}
\newcommand\ee{\end{equation}}

\allowdisplaybreaks[4]

\begin{document}

\tolerance=5000

\title{Reviving non-Minimal Horndeski-like Theories after GW170817: Kinetic Coupling Corrected Einstein-Gauss-Bonnet Inflation}
\author{V.K.~Oikonomou,$^{1,2,3}$\,\thanks{v.k.oikonomou1979@gmail.com}F.P.
Fronimos,$^{1}$\,\thanks{fotisfronimos@gmail.com}}
\affiliation{$^{1)}$ Department of Physics, Aristotle University
of Thessaloniki, Thessaloniki 54124,
Greece\\
$^{2)}$ Laboratory for Theoretical Cosmology, Tomsk State
University of Control Systems and Radioelectronics, 634050 Tomsk,
Russia (TUSUR)\\
$^{3)}$ Tomsk State Pedagogical University, 634061 Tomsk,
Russia\\}

\tolerance=5000

\begin{abstract}
After the recent GW170817 event of the two neutron stars merging,
many string corrected cosmological theories confronted the
non-viability peril. This was due to the fact that most of these
theories produce massive gravitons primordially. Among these
theories were the ones containing a non-minimal kinetic coupling
correction term in the Lagrangian. In this work we demonstrate how
these theories may be revived and we show how these theories can
produce primordial gravitational waves with speed $c_T^2=1$ in
natural units, thus complying with the GW170817 event. As we show,
if the gravitational action of an Einstein-Gauss-Bonnet theory
also contains a kinetic coupling of the form $\sim \xi(\phi)
G^{\mu\nu}\partial_\mu\phi\partial_\nu\phi$, the condition of
having primordial massless gravitons, or equivalently primordial
gravitational waves with speed $c_T^2=1$ in natural units, results
to certain conditions on the scalar field dependent coupling
function of the Gauss-Bonnet term, which is also the non-minimal
coupling of the kinetic coupling. We extensively study the
phenomenological implications of such a theory focusing on the
inflationary era, by only assuming slow-roll dynamics for the
scalar field. Accordingly, we briefly study the case that the
scalar field evolves in a constant-roll way. By using some
illustrative examples, we demonstrate that the viability of the
theoretical framework at hand may easily be achieved. Also,
theories containing terms of the form $\sim \xi(\phi)\Box\phi
g^{\mu\nu}\partial_\mu\phi\partial_\nu\phi$ and $\sim
\xi(\phi)\left(g^{\mu\nu}\partial_\mu\phi\partial_\nu\phi\right)^2$
are also lead to the same gravitational wave speed as the theory
we shall study in this paper, so this covers a larger class of
Horndeski theories.
\end{abstract}


\maketitle

\section{Introduction}

The primordial Universe is one of the mysteries that modern
theoretical cosmologists and physicists are trying to understand.
Our approach to the primordial Universe is through a regime of
classical gravity towards to the unknown era of quantum gravity or
the ``theory of everything'' which is believed to govern the small
scale physics and unifies all forces in nature. In between the
classical gravity regime and the quantum gravity regime,
theoretical cosmologists believe that an era of abrupt
acceleration occurred, known as inflationary era. This era is at
the border of classical and quantum gravity, so many believe that
it should have some imprints of the quantum gravity era, in the
effective theory that controls the inflationary regime. The most
appealing theoretical proposal for the ``theory of everything'' is
nowadays string theory in all its different forms. If string
theory controls the quantum gravity era, it is reasonable to
expect that the effective theory that controls the inflationary
era might have terms which originate in the complete M-theory
Lagrangian. In some sense the classical inflationary theory should
contain several string theory corrections. A particularly
appealing class of theories which contains string theory motivated
corrections is Einstein-Gauss-Bonnet theory of gravity
\cite{Hwang:2005hb,Nojiri:2006je,Cognola:2006sp,Nojiri:2005vv,Nojiri:2005jg,Satoh:2007gn,Bamba:2014zoa,Yi:2018gse,Guo:2009uk,Guo:2010jr,Jiang:2013gza,Kanti:2015pda,vandeBruck:2017voa,Kanti:1998jd,Kawai:1999pw,Nozari:2017rta,Chakraborty:2018scm,Odintsov:2018zhw,Kawai:1998ab,Yi:2018dhl,vandeBruck:2016xvt,Kleihaus:2019rbg,Bakopoulos:2019tvc,Maeda:2011zn,Bakopoulos:2020dfg,Ai:2020peo,Easther:1996yd,Antoniadis:1993jc,Antoniadis:1990uu,Odintsov:2020sqy,Odintsov:2020zkl,Odintsov:2019clh}.
These theories, along with several extensions containing kinetic
coupling corrections \cite{Hwang:2005hb} are known to provide a
viable inflationary era which can in many cases be compatible with
the observational data, see Refs.
\cite{Hwang:2005hb,Nojiri:2006je,Cognola:2006sp,Nojiri:2005vv,Nojiri:2005jg,Satoh:2007gn,Bamba:2014zoa,Yi:2018gse,Guo:2009uk,Guo:2010jr,Jiang:2013gza,Kanti:2015pda,vandeBruck:2017voa,Kanti:1998jd,Kawai:1999pw,Nozari:2017rta,Chakraborty:2018scm,Odintsov:2018zhw,Kawai:1998ab,Yi:2018dhl,vandeBruck:2016xvt,Kleihaus:2019rbg,Bakopoulos:2019tvc,Maeda:2011zn,Bakopoulos:2020dfg,Ai:2020peo,Easther:1996yd,Antoniadis:1993jc,Antoniadis:1990uu,Odintsov:2020sqy,Odintsov:2020zkl,Odintsov:2019clh}
and references therein for details.

However, Einstein-Gauss-Bonnet theories have a serious flaw,
namely the predict a non-zero graviton mass during the primordial
inflationary era. This was not a problem until recently, the
GW170817 event utterly changed the scenery in modern theoretical
cosmology, excluding many modified gravity theories from being
viable descriptions of the Universe, see Ref.
\cite{Ezquiaga:2017ekz} for a detailed account on this topic. This
is because the two neutron stars merging event GW170817
\cite{GBM:2017lvd} indicated that the gravitational waves arrived
almost simultaneously with the gamma rays emitted from the neutron
stars merging. Correspondingly, this observation clearly shows
that the gravitational wave speed $c_T$ is nearly equal to unity,
that is $c_T^2=1$ in natural units, which is equal to the speed of
light. Thus theories that do not predict a massless graviton
should no longer be considered as successful description of our
Universe at large or small scales. It is remarkable though that
many modified gravity theories
\cite{Nojiri:2017ncd,Nojiri:2010wj,Nojiri:2006ri,Capozziello:2011et,
Capozziello:2010zz,delaCruzDombriz:2012xy,Olmo:2011uz}, still
remain compatible with the observations, like $f(R)$ gravity or
Gauss-Bonnet theories.

Now the question is, why an event that is observed at small
redshifts should affect so seriously theories that predict a
primordial massive graviton? The answer to this question is
simple, there is no reason for the graviton to change its mass
during the evolution of the Universe. So the answer to the above
question, can also be cast in a question, why should the graviton
mass during the inflationary era be different from the gravitons
emitted from low redshift astrophysical sources? No particle
physics process can explain why the graviton should have different
mass during inflation, the post-inflationary era and at present
time, at least to our knowledge.

In view of this way of thinking, in a previous work we studied how
Einstein-Gauss-Bonnet theories can be compatible to the GW170817
event, and produce primordial gravitational waves with speed equal
to that of light
\cite{Odintsov:2020sqy,Odintsov:2020zkl,Odintsov:2019clh}. As we
showed, the condition $c_T^2=1$ restricts significantly the
functional forms of the Gauss-Bonnet scalar coupling function $\xi
(\phi)$ and the scalar potential. We also demonstrated that the
GW170817-compatible theories can produce a viable inflationary
era, compatible with the latest Planck data \cite{Akrami:2018odb}.

In this work, we extend the study performed in our previous work
\cite{Odintsov:2020sqy,Odintsov:2020zkl,Odintsov:2019clh},
including in the effective action of Einstein-Gauss-Bonnet
theories non-minimal kinetic coupling terms of the form $\sim
\xi(\phi) G^{\mu\nu}\partial_\mu\phi\partial_\nu\phi$. These type
of theories belong to the more general class of Horndeski theories
\cite{horndeskioriginal,Kobayashi:2019hrl,Kobayashi:2016xpl,Crisostomi:2016tcp,Bellini:2015xja,Gleyzes:2014qga,Lin:2014jga,Deffayet:2013lga,Bettoni:2013diz,Koyama:2013paa,Starobinsky:2016kua,Capozziello:2018gms,BenAchour:2016fzp,Starobinsky:2019xdp},
and are also known as kinetic coupling theories
\cite{Sushkov:2009hk,Minamitsuji:2013ura,Saridakis:2010mf,Barreira:2013jma,Sushkov:2012za,Barreira:2012kk,Skugoreva:2013ooa,Gubitosi:2011sg,Matsumoto:2015hua,Deffayet:2010qz,Granda:2010hb,Matsumoto:2017gnx,Gao:2010vr,Granda:2009fh,Germani:2010gm,Fu:2019ttf}
and after the GW170817 event, these theories were significantly
looked down upon, due to the fact that Horndeski theories produce
primordial gravitational waves with speed less than that of light.
Actually, Horndeski theories in view of GW170817 were also studied
and discussed in Refs.
\cite{Dima:2017pwp,Kreisch:2017uet,Arai:2017hxj}. In this paper,
we shall revive in a concrete way the Horndeski theories
containing kinetic couplings of the form $\sim \xi(\phi)
G^{\mu\nu}\partial_\mu\phi\partial_\nu\phi$, by imposing the
constraint that the gravitational wave speed is equal to that of
light. As we show in detail, the constraint $c_T^2=1$ results to a
differential equation that determines uniquely the dynamical
evolution of the scalar field. In addition, the coupling function
$\xi (\phi)$ and the scalar potential are not freely chosen, but
must satisfy a specific differential equation. Then by assuming
that the slow-roll conditions hold true, we show that the
inflationary phenomenology of GW170817-compatible kinetic coupling
corrected Einstein-Gauss-Bonnet theory can be compatible with the
latest Planck data \cite{Akrami:2018odb}. We also perform the
study in the case of a constant-roll evolution for the scalar
field, and we show that the viability can be achieved in this case
of dynamical evolution too. Finally, as we show, the same
constraint that renders the kinetic coupling corrected
Einstein-Gauss-Bonnet theories compatible with the GW170817 event,
also renders theories containing terms $\sim \xi(\phi)\Box\phi
g^{\mu\nu}\partial_\mu\phi\partial_\nu\phi$ and $\sim
\xi(\phi)\left(g^{\mu\nu}\partial_\mu\phi\partial_\nu\phi\right)^2$
also compatible with the GW170817 event. Thus a broader class of
Horndeski theories can be revived, however we restrict ourselves
to the simplest case of having only kinetic coupling terms,
because the extended theories are quite more difficult to study
with regard to their inflationary phenomenology.

\section{Non-minimal Kinetic Coupling Corrected Einstein-Gauss-Bonnet Inflationary Phenomenology: Essential Features and Compatibility with GW170817 Constraints}

In principle, Einstein-Gauss-Bonnet gravity is simply a string
corrected scalar field theory with minimal coupling, thus a
natural extension of this string corrected theory can be obtained
by simply adding a non-minimal kinetic coupling, or other higher
order string corrections containing derivatives of the scalar
field, see \cite{Hwang:2005hb} for more details on all the
possible string corrections. In this work we shall assume that the
gravitational action of a minimally coupled scalar field contains
two string corrections, namely a Gauss-Bonnet term with a scalar
field dependent coupling and a non-minimal kinetic coupling term,
and the action has the following form,
\begin{equation}
\centering \label{action}
\mathcal{S}=\int{d^4x\sqrt{-g}\left(\frac{R}{2\kappa^2}-\frac{\omega}{2}g^{\mu\nu}\partial_\mu\phi\partial_\nu\phi-V(\phi)-\xi(\phi)\mathcal{G}-c\xi(\phi)
G^{\mu\nu}\partial_\mu\phi\partial_\nu\phi\right)}\, ,
\end{equation}
where $R$ denotes as usual the Ricci scalar, $g$ is the
determinant of the metric tensor, $\kappa=\frac{1}{M_P}$ with
$M_P$ being the reduced Planck mass. Also $V(\phi)$ is the
potential of the scalar field, $\xi(\phi)$ is the Gauss-Bonnet
scalar coupling function which is coupled to the Gauss-Bonnet
topological invariant, with the latter being defined as
$\mathcal{G}=R_{\mu\nu\sigma\rho}R^{\mu\nu\sigma\rho}-4R_{\mu\nu}R^{\mu\nu}+R^2$,
and $R_{\mu\nu\sigma\rho}$ and $R_{\mu\nu}$ are the Riemann
curvature tensor and the Ricci tensor respectively. Finally, the
kinetic coupling term is the one $\sim c\xi(\phi)
G^{\mu\nu}\partial_\mu\phi\partial_\nu\phi$, where $G^{\mu\nu}$
stands for the Einstein tensor
$G^{\mu\nu}=R^{\mu\nu}-\frac{1}{2}g^{\mu\nu}R$, $c$ is a free
constant parameter with mass dimensions $[m]^{-2}$. Hence in this
work, we used two kind of string corrections for the minimally
coupled scalar field, the one used previously which is
$\xi(\phi)\mathcal{G}$ and the second type denoted as
$c\xi(\phi)G^{\mu\nu}\partial_\mu\phi\partial_\nu\phi$. Also, we
made use of $\omega$ multiplying the kinetic term of the scalar
field in the gravitational action, just for generality and in
order to provide expression in the following sections which are as
general as possible. In the following we shall take $\omega=1$ in
order to have a canonical kinetic term for the scalar field.

For the purposes of studying inflationary dynamics in this paper,
we shall assume that the gravitational background will be that of
a flat Friedman-Robertson-Walker (FRW) with line element,
\begin{equation}
\centering
\label{metric}
ds^2=-dt^2+a^2\delta_{ij}dx^idx^j\, ,
\end{equation}
where $a(t)$ denotes as usual the scale factor. This choice for
the line element is also convenient in order to obtain simple
functional forms for the curvature related terms, since now both
the Ricci scalar and the Gauss-Bonnet topological invariant can be
written in terms of Hubble's parameter $H=\frac{\dot a}{a}$ as
$R=6(2H^2+\dot H)$ and $\mathcal{G}=24H^2(\dot H+H^2)$, where the
``dot'' as usual implies differentiation with respect to cosmic
time $t$. Also we shall assume that the scalar field is
homogeneous, so it has a time dependence solely.

At this point let us investigate the implication of the
requirement that the speed of the primordial gravitational waves
is equal to unity, that is $c_T^2=1$ in natural units, for the
theory of the action (\ref{action}). As it was shown in Ref.
\cite{Hwang:2005hb}, the speed of the tensor mode of the
primordial curvature perturbations for the action (\ref{action})
is equal to,
\begin{equation}
\centering
\label{cT}
c_T^2=1-\frac{Q_f}{2Q_t}\, ,
\end{equation}
where in the case at hand, the term $Q_f$ is equal to,
\begin{equation}\label{extraqf}
Q_f=16(\ddot\xi-H\dot\xi)+4c\xi\dot\phi^2\, ,
\end{equation}
and in addition $Q_t=\frac{1}{\kappa^2}-8\dot\xi
H+c\xi\dot\phi^2$. Due to the presence of the extra kinetic
coupling string correction, the functional forms of the auxiliary
parameters $Q_f$ and $Q_t$ are intrinsically different in the case
at hand, in comparison to the usual Einstein-Gauss-Bonnet theory,
as it can be seen by comparing the above with the ones we used in
Ref. \cite{Odintsov:2020sqy}. As it is obvious from Eq.
(\ref{extraqf}), the compatibility with the GW170817 event can be
once again restored by demanding that $Q_f=0$ or at least
$Q_f\simeq 0$, which in view of the functional form of $Q_f$ given
in Eq. (\ref{extraqf}) results to the following constraint,
\begin{equation}
\centering \label{constraint}
4(\ddot\xi-H\dot\xi)+c\xi\dot\phi^2=0\, .
\end{equation}
It is obvious that for the case $c=0$, one obtains again the case
studied in \cite{Odintsov:2020sqy}, so the usual
Einstein-Gauss-Bonnet theory. We can rewrite this expression in
terms of the scalar field $\phi$ and by using
$\frac{d}{dt}=\dot\phi\frac{d}{d\phi}$ then the constraint
(\ref{constraint}) can be rewritten as follows,
\begin{equation}
\centering
\label{difeq}
4(\xi''\dot\phi^2+\xi'\ddot\phi-H\xi'\dot\phi)+c\xi\dot\phi^2=0\, ,
\end{equation}
where the ``prime'' denotes differentiation with respect to the
scalar field, and this notation will be kept for the rest of this
paper. It is worth mentioning that the exactly same constraint
would result in order for the gravitational wave speed to be equal
to unity, even if the action contained additional string
corrections of the form $c_1\xi(\phi)\Box\phi
g^{\mu\nu}\partial_\mu\phi\partial_\nu\phi$ and
$c_2\xi(\phi)\left(g^{\mu\nu}\partial_\mu\phi\partial_\nu\phi\right)^2$
\cite{Hwang:2005hb}, however this case would possibly lead to
highly complicated and difficult to tackle analytically equations
of motion, so we do not consider these in this paper.

Since we are aiming in studying the inflationary era, we shall
assume that the slow-roll conditions hold true, that is
$\dot{H}\ll H^2$ and also that the slow-roll conditions hold true
for the scalar field, that is $\ddot\phi\ll H\dot\phi$. In effect,
we also have $\xi'\ddot\phi\ll \xi'H\dot\phi$. Thus, the
differential equation (\ref{difeq}) can be simplified and it
reads,
\begin{equation}
\centering
\label{dotphi}
\dot\phi=\frac{4H\xi'}{4\xi''+c\xi}\, ,
\end{equation}
It is important to make certain comments on this result. Firstly,
it is reminiscing of the one previously acquired result in Ref.
\cite{Odintsov:2020sqy}, which corresponds to the case of having
$c=0$, with the only difference being a factor $4$, but this is a
feature which was expected. Also, owning to the fact that the time
derivative of the scalar field is written now proportionally to
Hubble's parameter and the Gauss-Bonnet coupling function, the
same principles apply. In particular, all the functions can be
written in terms of the scalar field instead of cosmic time $t$
and furthermore, the scalar functions of the model are
interconnected. Particularly, the derivative of the scalar field
$\dot{\phi}$ must satisfy both Eq. (\ref{dotphi}) and the equation
of motion of the scalar field. The existence of $c$ in the
denominator makes it also a bit difficult to simplify the
expression of Eq. (\ref{dotphi}), hence the strategy we followed
in Ref. \cite{Odintsov:2020sqy} in which case the choices of the
coupling function $\xi(\phi )$ were done in such a way so that the
ratio $\xi'/\xi''$ gets simplified, cannot be used in the present
paper.

Now let us obtain the field equations for the gravitational action
(\ref{action}), by varying the action with respect to the metric
and with respect to the scalar field. By doing so, one obtains the
field equations for gravity, which in our case deviate from the
Einstein field equations, and the equations of motion read,
\begin{equation}
\centering \label{motion1}
\frac{3H^2}{\kappa^2}=\frac{1}{2}\dot\phi^2+V+24\dot\xi H^3-9c\xi
H^2\dot\phi^2\, ,
\end{equation}
\begin{equation}
\centering \label{motion2} -\frac{2\dot
H}{\kappa^2}=\dot\phi^2-8H^2(\ddot\xi-H\dot\xi)-16\dot\xi H\dot
H+c\left(2\xi(\dot
H-3H^2)\dot\phi^2+4H\xi\dot\phi\ddot\phi+2H\dot\xi\dot\phi^2\right)\,
,
\end{equation}
\begin{equation}
\centering \label{motion3} V'+
(\ddot\phi+3H\dot\phi)+\xi'\mathcal{G}-3c\left(H^2(\dot\xi\dot
H+2\xi\ddot\phi)+2H(2\dot H+3H^2)\xi\dot\phi\right)=0\, .
\end{equation}
The third equation stands for the differential equation of the scalar
potential due to the fact that Eq. (\ref{dotphi}) was derived from
the realization that the graviton must be massless during
the inflationary era. It is the general expression with zero approximations
assumed.
As it is obvious, in the case at hand, the field equations
corresponding to the action (\ref{action}) have a lengthy
functional form. Also the system of equations of motion is very
complex and an analytical solution cannot be extracted easily.
However, we shall try and simplify the equations above in order to
obtain a viable phenomenology. In the following, we shall make two
types of assumptions. Firstly, as it was mentioned previously, the
slow-roll approximations for the Hubble rate and the scalar field
will be implemented and in particular, we shall assume that,
\begin{align}
\centering \label{slowroll} \dot H&\ll H^2 &\frac{1}{2}
\dot\phi^2&\ll V& \ddot\phi&\ll H\dot\phi\, .
\end{align}
Furthermore we shall assume that the string corrective terms
are also negligible from the equations of motion, leaving only
the canonical scalar field terms. In order to not get repetitive
we showcase the aforementioned approximation for the first equation
of motion, which is
\begin{equation}\label{extraconditions}
24\dot\xi H^3\ll V,\,\,\,-9c\xi H^2\dot\phi^2\ll V\, ,
\end{equation}
but in principle, we demand that $\xi\ll1$ and $c\xi\ll1$ with
all the possible combinations appearing in equations (\ref{motion1})
through (\ref{motion3}). Obviously, in the end we must explicitly
check whether these conditions are satisfied for all the
phenomenologically viable models we shall study. In view of the
above assumptions, the equations of motion are greatly simplified
and these read,
\begin{equation}
\centering
\label{motion4}
H^2=\frac{\kappa^2V}{3}\, ,
\end{equation}
\begin{equation}
\centering \label{motion5} \dot H=-\frac{\kappa^2
H^2}{2}\left(\frac{4\xi'}{4\xi''+c\xi}\right)^2\, ,
\end{equation}
\begin{equation}
\centering \label{motion6} V'+3  H^2\frac{4\xi'}{4\xi''+c\xi}=0\,
,
\end{equation}
where in the last equation, we made use of Eq. (\ref{dotphi}). The
above equations of motion are simpler in comparison to those
obtained without the slow-roll assumption, and thus in the case at
hand analytic results can be obtained and the phenomenology of the
non-minimal kinetic coupling corrected Einstein-Gauss-Bonnet
theory can be investigated in a simpler and direct way. The
advantage of the slow-roll assumption in the case at hand is that,
given the scalar coupling function $\xi(\phi)$, the scalar
potential can be determined directly by using the equations of
motion, as we show in the illustrative examples we chose to
present in the next section. It is important to note again
however, something that we already mentioned earlier in this
section, that is, in the end it is vital to check whether the
approximations imposed hold true, for each model that yields a
viable phenomenology. It is conceivable that a model that yields a
viable phenomenology, but still fails to satisfy the initial
assumptions, is intrinsically wrong.

As it will be apparent shortly, the scalar coupling $\xi(\phi)$
plays a crucial role in the theory at hand, and critically affects
the slow-roll indices and the resulting observational indices,
namely the spectral index of the primordial scalar perturbations,
and the tensor-to-scalar ratio. Also as it will be shown shortly,
an elegant choice of $\xi(\phi)$ will results to quite elegant
final expressions of the aforementioned phenomenology related
parameters.

Now using the slow-roll assumptions we shall derive the analytic
expressions for the slow-roll indices for the theory at hand. For
the non-minimal kinetic coupling corrected Einstein-Gauss-Bonnet
theory the slow-roll indices have the general expressions
\cite{Hwang:2005hb},
\begin{align}
\centering \label{indices} \epsilon_1&=-\frac{\dot
H}{H^2}&\epsilon_2&=\frac{\ddot\phi}{H\dot\phi}&\epsilon_3&=0&\epsilon_4&=\frac{\dot
E}{2HE}&\epsilon_5&=\frac{Q_a}{2HQ_t}&\epsilon_6&=\frac{\dot
Q_t}{2HQ_t}\, ,
\end{align}
where $Q_a=-8\dot\xi H^2+4c\xi H\dot\phi^2$,
$E=\frac{1}{(\kappa\dot\phi)^2}\left(
\dot\phi^2+\frac{3Q_a^2}{2Q_t}+Q_c\right)$ and
$Q_c=-6c\xi\dot\phi^2H^2$. Thus, using the slow-roll equations of
motion, namely Eqs. (\ref{motion4})-(\ref{motion6}), we obtain,
\begin{equation}
\centering \label{index1} \epsilon_1=\frac{\kappa^2
}{2}\left(\frac{4\xi'}{4\xi''+c\xi}\right)^2\, ,
\end{equation}
\begin{equation}
\centering
\label{index2}
\epsilon_2=\frac{4\xi''}{4\xi''+c\xi}-\epsilon_1-\frac{4\xi'(4\xi'''+c\xi')}{(4\xi''+c\xi)^2}\, ,
\end{equation}
\begin{equation}
\centering
\label{index4}
\epsilon_4=\frac{2\xi'}{4\xi''+c\xi}\frac{E'}{E}\, ,
\end{equation}
\begin{equation}
\centering
\label{index5}
\epsilon_5=\frac{-16\frac{\xi'^2H^2}{4\xi''+c\xi}+4c\xi H^2\left(\frac{4\xi'}{4\xi''+c\xi}\right)^2}{\frac{1}{\kappa^2}-32\frac{\xi'^2H^2}{4\xi''+c\xi}+32c\xi H^2\left(\frac{\xi'}{4\xi''+c\xi}\right)^2}\, ,
\end{equation}
\begin{equation}
\centering
\label{index6}
\epsilon_6=-\frac{4H^2\frac{4\xi'^2}{4\xi''+c\xi}(\frac{4\xi''}{4\xi''+c\xi}+\epsilon_2-\epsilon_1)+cH^2\left(\frac{4\xi'}{4\xi''+c\xi}\right)^2\left(\frac{2\xi'^2}{4\xi''+c\xi}+\xi\epsilon_2\right)}{\frac{1}{\kappa^2}-32\frac{\xi'^2}{4\xi''+c\xi}H^2+c\xi H^2\left(\frac{4\xi'}{4\xi''+c\xi}\right)^2}\, ,
\end{equation}
and the auxiliary parameters $Q_a$, $Q_t$, $Q_d$, $Q_e$ in turn
take the following form,
\begin{equation}
\centering
\label{Qa}
Q_a=-32\frac{\xi'^2}{4\xi''+c\xi}H^3+4cH^3\xi\left(\frac{4\xi'}{4\xi''+c\xi}\right)^2\, ,
\end{equation}
\begin{equation}
\centering
\label{Qt}
Q_t=\frac{1}{\kappa^2}-32\frac{\xi'^2}{4\xi''+c\xi}H^2+c\xi H^2\left(\frac{4\xi'}{4\xi''+c\xi}\right)^2\, ,
\end{equation}
\begin{equation}
\centering
\label{Q_c}
Q_c=-6c\xi H^4\left(\frac{4\xi'}{4\xi''+c\xi}\right)^2\, ,
\end{equation}
\begin{equation}
\centering
\label{Q_d}
Q_d=-4c\xi H^2\left(\frac{4\xi'}{4\xi''+c\xi}\right)^2\dot H\, ,
\end{equation}
\begin{equation}
\centering \label{Qe} Q_e=-32\frac{4\xi'^2H}{4\xi''+c\xi}\dot
H+16cH\frac{\xi'}{4\xi''+c\xi}\left(\xi'H^2\left(\frac{4\xi'}{4\xi''+c\xi}\right)^2+2\xi\ddot\phi-8\xi
H^2\frac{\xi'}{4\xi''+c\xi}\right)\, ,
\end{equation}
where in order not to make $Q_e$ lengthy, the $\ddot\phi$ is left
as it is but essentially it is equal to $\ddot\phi=\epsilon_2
H\dot\phi=4\epsilon_2 H^2\frac{\xi'}{4\xi''+c\xi}$. The parameter
$Q_e$ is not used here but is introduced now for convenience since
it shall be utilized in the subsequent calculations. It is
important to make two comments here. Firstly, the indices
$\epsilon_5$ and $\epsilon_6$ have quite different forms thus it
is not expected that their values will be similar. This is
attributed to the non-minimal kinetic coupling term proportional
to $c$. Also, for $c=0$, we obtain the same equations as in Ref.
\cite{Odintsov:2020sqy}. In the context of the present formalism,
only the first two slow-roll indices have simple functional forms
whereas the rest are quite complicated, especially the index
$\epsilon_4$ which at best its form could be characterized as
lengthy. Nonetheless, the slow-roll indices are of paramount
importance since in this approach, the observational indices,
namely the scalar spectral index of primordial curvature
perturbations $n_S$, the tensor spectral index $n_T$ and the
tensor-to-scalar ratio $r$ are given in terms of the slow-roll
indices as follows \cite{Hwang:2005hb},
\begin{align}
\centering \label{observed}
n_S&=1-2\frac{2\epsilon_1+\epsilon_2+\epsilon_4}{1-\epsilon_1}&n_T&=-2\frac{\epsilon_1+\epsilon_6}{1-\epsilon_1}&r&=16\left|\left(\epsilon_1+\frac{\kappa^2(2Q_c+Q_d-HQ_e)}{4H^2}\right)\frac{c_A^3}{\kappa^2Q_t}\right|\,
,
\end{align}
where, $c_A$ denotes the sound wave velocity and it is given by
the formula,
\begin{equation}
\centering \label{cA} c_A^2=1+\frac{2Q_tQ_d+Q_aQ_e}{2
\dot\phi^2Q_t+3Q_a^2+2Q_cQ_t}\, .
\end{equation}
Also it is important to note that the formulas quoted in Eq.
(\ref{observed}) hold true only if the slow-roll assumptions hold
true for the slow-roll indices, that is when $\epsilon_i\ll 1$,
and a refined derivation of the formulas (\ref{observed}) was
given in Ref. \cite{Oikonomou:2020krq}. Lastly, a vital ingredient
of a theory that is analytically tractable, is the functional form
of the $e$-foldings number, which shall be used in the subsequent
calculations. Since $N=\int_{t_i}^{t_f}{Hdt}$ where the time
difference $t_f-t_i$ signifies the duration of the inflationary
era and recalling Eq. (\ref{dotphi}), one obtains the following
result,
\begin{equation}
\centering
\label{efolds}
N=\int_{\phi_i}^{\phi_f}{\frac{4\xi''+c\xi}{4\xi'}d\phi}\, ,
\end{equation}
It is clear that the overall phenomenology acquired for the
non-minimal kinetic coupling corrected Einstein-Gauss-Bonnet
theory is not fundamentally different from the one obtained from
the simple Einstein-Gauss-Bonnet theory. In fact, many, if not
every parameter here is similar to the corresponding ones in Ref.
\cite{Odintsov:2020sqy}. Hence, it is expected that the same steps
will yield appealing characteristics, if not viability.
Specifically, since each object is written in terms of the scalar
field $\phi$, we shall evaluate its value during the initial and
final stage of the inflationary era. The latter can be extracted
by letting the first slow-roll index $\epsilon_1$ become of order
$\mathcal{O}(1)$. Afterwards, the initial value of the scalar
field $\phi_i$ at the first horizon crossing can be obtained
easily from the $e$-foldings number, hence this is the reason why
we rewritten it in terms of the scalar field. In the following
section, we shall extensively study the phenomenological viability
of several appropriately chosen models and when possible, we shall
directly compare the non-minimal kinetic coupling
Einstein-Gauss-Bonnet theory with the simple Einstein-Gauss-Bonnet
theory.

Before we proceed, it is worth making a statement. The condition that
$\xi(\phi)\ll1$ is not only imposed so as to get rid of extra terms which
are difficult to compute in the equations of motion nor is imposed so as
to get explicitly similar equations with the canonical scalar field case. In
fact, there exists an underlying reason which forces once to to accept that
the combination of $c\xi$ is many orders lesser than unity, and that is the
so called Jeans instabilities. Indeed, by quickly reviewing the numerical value
of the field propagation velocity $c_A$ introduced in Eq. (\ref{cA}), one clearly
sees that if $c$ and $\xi$ are left unchecked, neither compatibility with the Planck 2018
collaboration data can be achieved nor a smooth description of the inflationary
era. Essentially, $c_A$ can obtain either exaggerating superluminal values of
become complex with the imaginary part being non negligible. We shall refer to
this extensively in the following two examples but it is worth presenting such statement
at this point since a small value of $\xi$ clearly results in a gravitational action which
is quite close to Einstein's gravity but nonetheless contains important new information
about the dynamics of the universe.

\section{Confronting the GW170817-compatible Non-minimal Kinetic Coupling Corrected Einstein-Gauss-Bonnet With the Planck Data}

In this section we shall investigate several models that may
provide a viable phenomenology, by appropriately choosing the
scalar coupling function $\xi(\phi)$. The procedure we shall
follow is to choose and fix the Gauss-Bonnet scalar coupling
function firstly $\xi(\phi)$, accordingly we shall find the
corresponding scalar potential, and afterwards we derive the
corresponding expression of the scalar field from Eq.
(\ref{motion6}). Upon finding its expression, we shall demonstrate
how the slow-roll indices are influenced by such selection and
finally, by utilizing the form of slow-roll index $\epsilon_1$ and
the $e$-foldings number, the form of the scalar field during the
final stage of inflation and the first horizon crossing
respectively can be extracted. Hence, inserting the initial value
of the scalar field as an input in Eq. (\ref{observed}), the
observational indices can be numerically evaluated. In principle
the form of the scalar field during the horizon crossing depends
on the free parameters of the model used, hence before proceeding,
we shall choose appropriate values for these parameters. Thus, by
comparing the numerical value of the observational indices
obtained by the model, namely the scalar spectral index $n_S$, the
tensor spectral index $n_T$ and the tensor-to-scalar ratio $r$,
with the ones coming from the latest Planck 2018 collaboration,
the validity of the model can be ascertained. As a last step, it
is crucial to examine the validity of all the approximations
imposed in the previous section. Let us now proceed with the
examples.

\subsection{The Choice Of A Trigonometric Scalar Coupling $\xi(\phi)$}

\begin{figure}[h!]
\centering
\label{plot1}
\includegraphics[width=17pc]{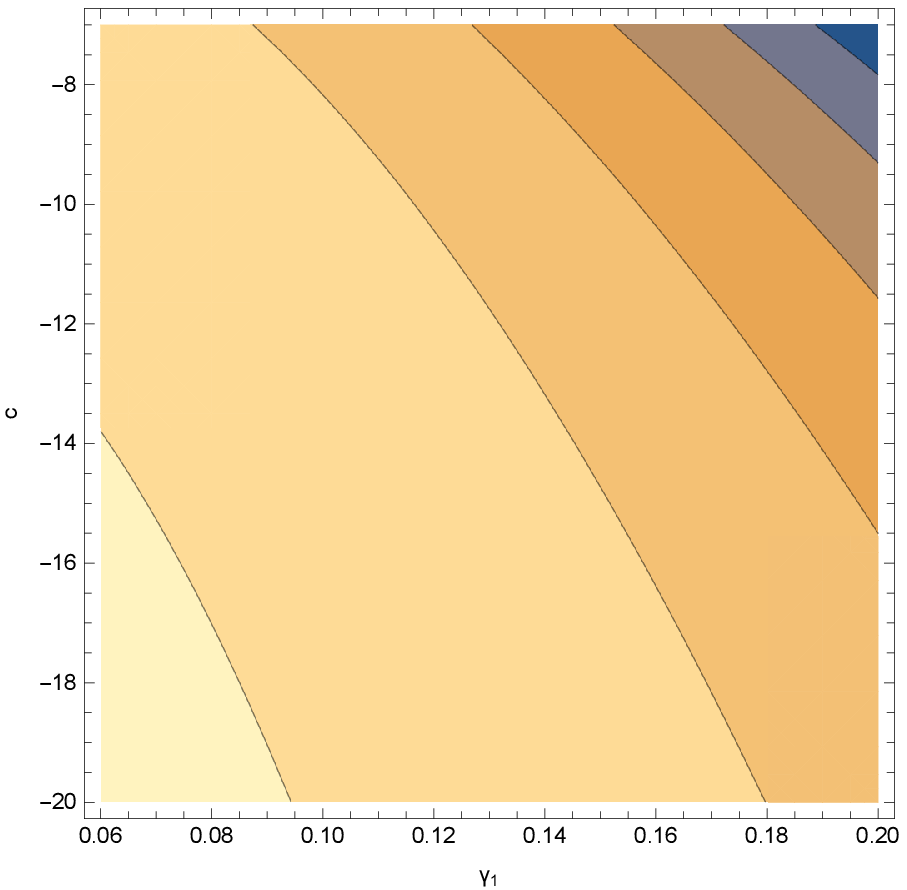}
\includegraphics[width=2.5pc]{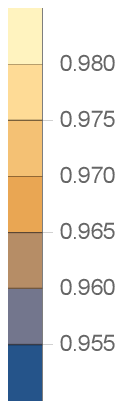}
\includegraphics[width=17pc]{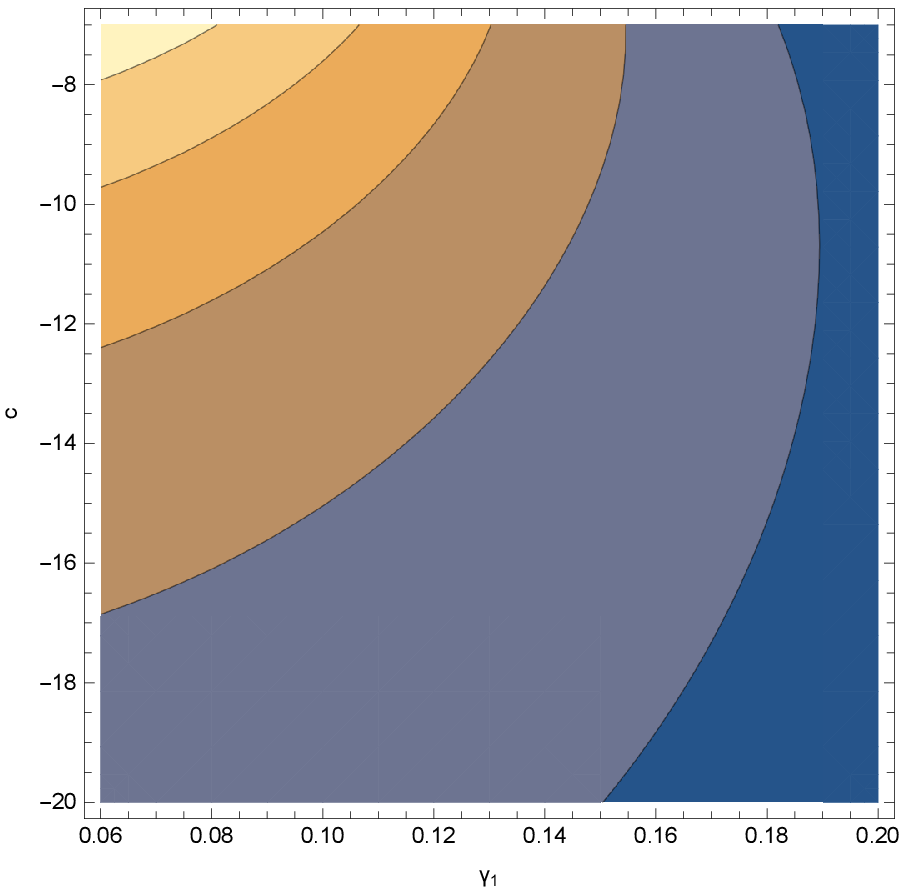}
\includegraphics[width=3pc]{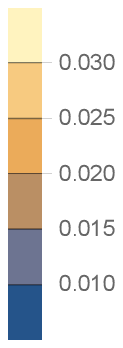}
\caption{ The spectral index of primordial curvature perturbations
$n_S$ (left) and tensor-to-scalar ratio $r$ (right) as functions
of $c$ and $\gamma_1$, ranging from [-20,-7] and [0.06,0.2]
respectively. It can easily be inferred that the phenomenological
viability of the model can be achieved for a wide range of the
free parameters.}
\end{figure}

We commence our study by assuming the Gauss-Bonnet has the
following form,
\begin{equation}
\centering
\label{xiA}
\xi(\phi)=\lambda_1\sin(\gamma_1\kappa\phi)\, ,
\end{equation}
where $\lambda_1$ and $\gamma_1$ are the free parameters of the
model. It can easily be inferred that this choice is somewhat
beneficial since $\xi''=-\gamma_1^2\kappa^2\xi$, therefore the
ratio $\frac{4\xi'}{4\xi''+c\xi}$ can be simplified. Consequently,
the corresponding scalar potential reads,
\begin{equation}
\centering \label{VA}
V(\phi)=V_1\sin(\gamma_1\kappa\phi)^{\frac{4\kappa^2
}{(2\gamma_1\kappa)^2-c}}\, ,
\end{equation}
where $c$ is the integration constant with mass dimensions $[
m]^{4}$ for consistency. In this case, we can see that the scalar
potential is also trigonometric as the Gauss-Bonnet Coupling is.
However, the potential has also an exponent which is not
necessarily an integer. Indeed, in the following we shall see that
the exponent obtains a value which is not an integer. Concerning
the first slow-roll index, it takes the following form,
\begin{equation}
\centering \label{index1A} \epsilon_1=2
\left(\frac{2\gamma_1\kappa^2\cot(\gamma_1\kappa\phi)}{c-(2\gamma_1\kappa)^2}\right)^2\,
,
\end{equation}
while we refrained from quoting the rest of the slow-roll indices
due to their lengthy form. Accordingly, the initial and final
value of the scalar field during the inflationary era are,
\begin{equation}
\centering
\label{phiiA}
\phi_i= \frac{1}{\gamma_1\kappa}\cos^{-1}\left(e^{-\frac{4N(\gamma_1\kappa)^2}{(2\gamma_1\kappa)^2-c}}\cos(\gamma_1\kappa\phi_f)\right)\, ,
\end{equation}
\begin{equation}
\centering \label{phifA}
\phi_f=\frac{1}{\gamma_1\kappa}\cot^{-1}\left(\frac{1}{\sqrt{2
}}\frac{(2\gamma_1\kappa)^2-c}{2\gamma_1\kappa^2}\right)\, ,
\end{equation}
with $\phi_f$ being obtained by equating $\epsilon_1(\phi_f)=1$.
Also $\phi_i$ is obtained by performing the integration in the
$e$-foldings number (\ref{efolds}) and substituting (\ref{phifA}),
and then solving the algebraic equation with respect to $\phi_i$.
So far, the procedure seems smooth while some of the slow-roll
indices which participate in the observed indices are perplexed.
Now let us proceed to the phenomenology of the present model. As
it can be seen, the viability of the model comes easily, since it
can be achieved for a wide range of the free parameters. For
example by assigning the following values to the free parameters
in reduced Planck units (where $\kappa=1$) ($\lambda_1$, $V_1$,
$N$, $\gamma_1$, $c$)=($10^{-8}$, $10^3$, 60, 0.15, -7) then the
observational indices are compatible with the latest Planck
constraints \cite{Akrami:2018odb}. In particular, the scalar
spectral index of primordial curvature perturbations obtains the
value $n_S=0.96554$, the tensor spectral index is equal to
$n_T=-0.0019893$ and finally the tensor-to-scalar ratio becomes
$r=0.015899$ which are obviously accepted values. Finally the
numerical values of the slow-roll indices are
$\epsilon_1=0.0009936$, $\epsilon_2=0.015228$,
$\epsilon_4=-4.05\cdot10^{-7}$, $\epsilon_5=-3.78\cdot10^{-8}$ and
finally $\epsilon_6=5\cdot10^{-10}$ which essentially is zero.
These values are indicative of the fact that the slow-roll
conditions implemented previously in the equations of motion are
indeed valid. Lastly, the sound wave velocity is also equal to
unity in natural units therefore the model is free of ghosts. The
previous designation might seem bizarre but was chosen since it
led to a viable phenomenology but as we also show, the assumptions
and approximations made in the previous sections are also
satisfied. Also in Fig. \ref{plot1} we can see that the model is
phenomenologically viable for a wide range of the free parameters.
The difference between the $\xi\mathcal{G}$ and $c\xi
G^{\mu\nu}\partial_\mu\phi\partial_\nu\phi$ seems to be the
constants and in particular the constant $\lambda_1$ of the
Gauss-Bonnet coupling and $c$. Our analysis showed that the
parameter $\lambda_1$ must be orders of magnitude lesser than
unity in order to achieve a phenomenologically acceptable
inflationary era. However, this can be attributed to the value of
$c$ which is trivial in this case. In addition, the parameters
$\lambda_1$ and $V_1$ seem to affect drastically the scalar
spectral index. This was also the case for $\gamma_1$ being
$0.15$ since it was the only value capable of doing so when
$\lambda_1$ and $V_1$ were fixed in these values, along with $c$,
so as not to violate the approximations made. Specifically, $c$
must be negative otherwise the scalar field obtains a complex
value during the first horizon crossing. Finally, the exponent of
the scalar potential is $0.5642$, which as stated before is not an
integer. Perhaps there exists a different designation for the free
parameters of the model at hand, which is capable of not only
producing an integer exponent but also produced compatible results
while simultaneously the approximated equations of motion are
valid. This case was not further studied.

An important comment that should be addressed here is the
numerical value of the tensor spectral index $n_T$. Since the
model was constrained so as to neglect string corrective terms,
however their impact is important since they participate in both
$\dot\phi$ and $V(\phi)$, it stands to reason that slow-roll index
$\epsilon_6$, which is comprised of such corrections, is inferior
to the first slow-roll index. This can easily be ascertained from
the previous numerical values of all slow-roll indices presented
above. As a result, the tensor spectral index can be reduced to
its usual form
\begin{equation}
\centering
\label{reducednT}
n_T=-2\epsilon_1\, ,
\end{equation}
and recalling the approximate form of $\epsilon_1$ in (\ref{index1})
it becomes apparent that the tensor spectral index is blue shifted,
meaning always negative. The only reason to obtain a red shifted index
is to let $\epsilon_6$, become dominant and of course negative. The latter
is not a possibility that can be ruled out however the first is out of the
question in this formalism, given that string corrective terms where assumed
to be inferior to $\dot\phi$ and $V(\phi)$, even though both quantities are
comprised of them. Even absent, string corrections have a strong impact on
the overall phenomenology indirectly.

Finally, we address the validity of the approximations made in the
previous section. Concerning the slow-roll indices, it was
mentioned some paragraphs above, that these respect the slow-roll
conditions. Also using for example the values ($\lambda_1$, $V_1$,
$N$, $\gamma_1$, $c$)=($10^{-8}$, $10^3$, 60, 0.15, -7), we have
$\dot H\sim\mathcal{O}(10^{-1})$ whereas
$H^2\sim\mathcal{O}(10^2)$, so the condition $\dot{H}\ll H^2$
holds true. In addition, $\frac{1}{2}
\dot\phi^2\sim\mathcal{O}(10^{-1})$ and similarly
$V\sim\mathcal{O}(10^2)$, so the assumption $\frac{1}{2}
\dot\phi^2\ll V$ holds true. Finally,
$\ddot\phi\sim\mathcal{O}(10^{-1})$ compared to
$H\dot\phi\sim\mathcal{O}(10)$, so the condition $\ddot\phi\ll
H\dot{\phi}$ holds true. Now let us see whether the rest of the
assumptions made in the previous sections indeed hold true. In
particular, for the first equation of motion Eq. (\ref{motion1}),
we have $24\dot\xi H^3\sim\mathcal{O}(10^{-5})$ and
$9c\xi\dot\phi^2H^2\sim\mathcal{O}(10^{-4})$ compared to
$V\sim\mathcal{O}(100)$, so the first condition in Eq.
(\ref{extraconditions}), namely $24\dot\xi H^3\ll V$ holds true.
Also, we have $16\dot\xi H\dot H\sim\mathcal{O}(10^{-8})$,
$8(\ddot\xi-H\dot\xi)\sim\mathcal{O}(10^{-5})$ and $c\left(2\xi
\dot\phi^2\dot
H-6\xi\dot\phi^2H^2+4H\xi\dot\phi\ddot\phi+2H\dot\xi\dot\phi^2\right)\sim\mathcal{O}(10^{-5})$
while $ \dot\phi^2\sim\mathcal{O}(10^{-1})$ hence Eq.
(\ref{motion5}) is indeed valid. Lastly, with regard to
(\ref{motion3}), we have
$\xi'\mathcal{G}\sim\mathcal{O}(10^{-3})$,
$3c\left(H^2(\dot\xi\dot H+2\xi\ddot\phi)+2H(2\dot
H+3H^2)\xi\dot\phi\right)\sim\mathcal{O}(10^{-3})$ while
$V'\sim\mathcal{O}(10)$. As a result, all the approximations made
are valid.

Let us now return to the case of the numerical value of $c\xi$.
As mentioned before, such combination is forced to take small
values otherwise the model becomes unstable. Indeed, by assuming
that $\lambda_1$ and $c$ are of order $\mathcal{O}(10^3)$,
 compatibility with recent observations vanishes since now the
scalar spectral index obtains an insane value of $n_S=3.14739$
and also $c_A=34$ at the end of the inflationary era, which means
that causality is violated. Such a problem however is not present
when the previous parameters take small values. This result is
at variance with the the Gauss-Bonnet case of $\xi(\phi)\mathcal{G}$
since now $\lambda_1$ alters the results significantly.

\begin{figure}[h!]
\centering
\label{plot2}
\includegraphics[width=17pc]{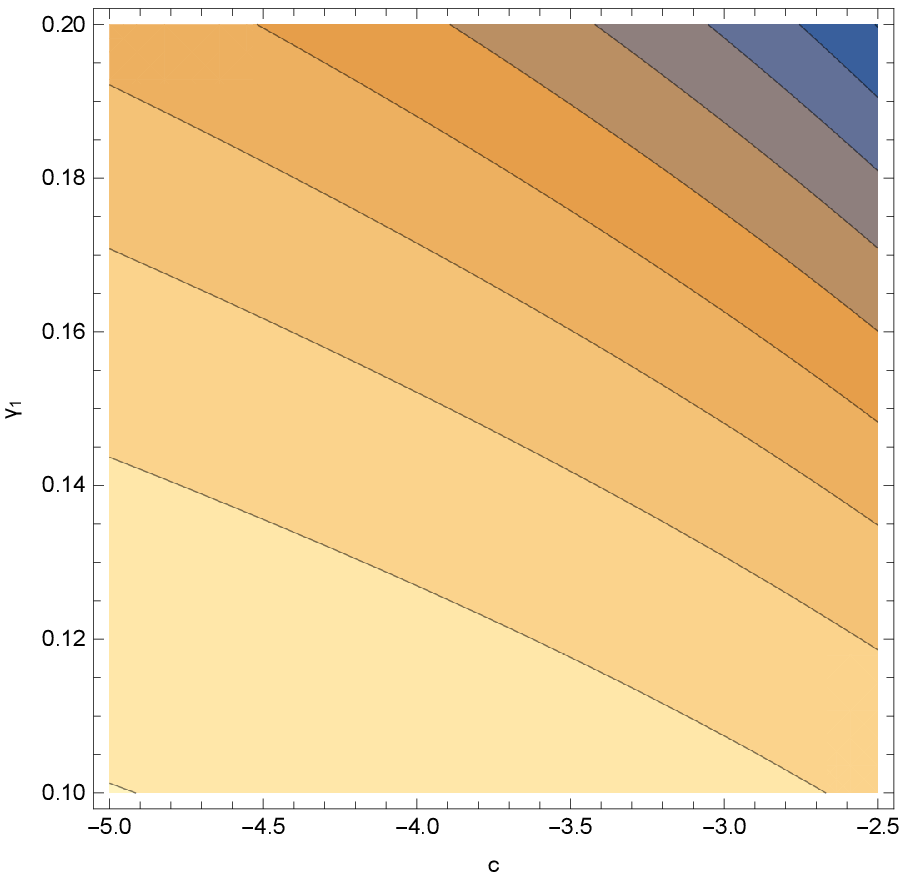}
\includegraphics[width=2.5pc]{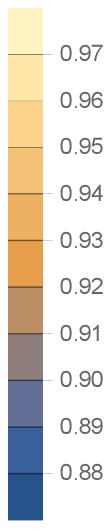}
\includegraphics[width=17pc]{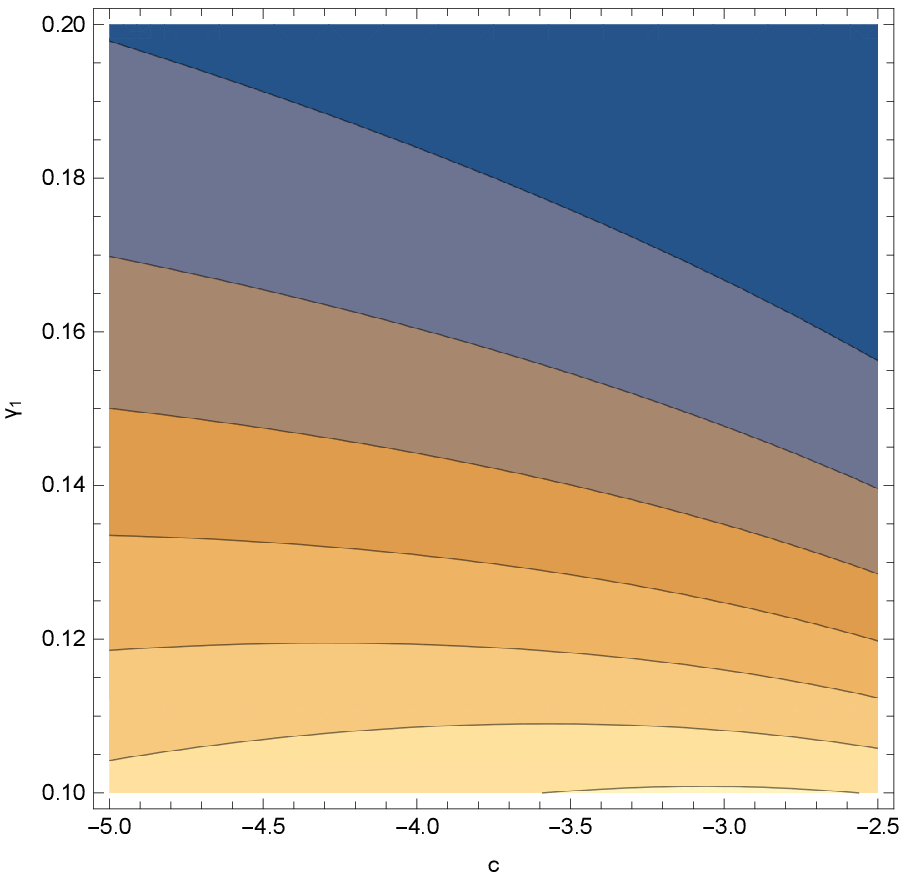}
\includegraphics[width=2.5pc]{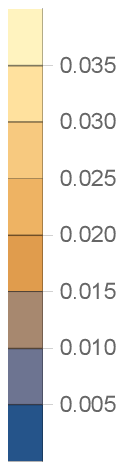}
\caption{Scalar spectral index $n_S$ (left) and tensor-to-scalar
ratio $r$ (right) depending on $c$ and $\gamma_1$, ranging from
[-5,-2.5] and [0.1,0.2] respectively for the case of cosine
Gauss-Bonnet coupling. }
\end{figure}

As a last comment, it is worth mentioning that the choice for a
cosine as a Gauss-Bonnet coupling function is also capable of
producing a viable phenomenology. In particular, assuming that
$\xi(\phi)=\lambda_1\cos(\gamma_1\kappa\phi)$ then by altering
only $\gamma_1$ and $c$ to $c=-3$ and $\gamma_1=0.1$ manages to
produce viable observed indices as $n_S=0.96239$, $r=0.0355884$
and $n_T=-0.00445847$ while simultaneously the sound wave velocity
is once again equal to unity. In Fig. \ref{plot2} we present the
dependence of the observational indices on the same free
parameters, namely  $c$ and $\gamma_1$. Moreover, all the
approximations made in the previous section, are satisfied once
again. It becomes apparent that the same phenomenology is gained
in the case of a cosine as well. Thus, one could argue that the
general Gauss-Bonnet scalar coupling function reads
$\xi(\phi)=\lambda_1\sin(\gamma_1\kappa\phi+\theta)$ where
$\theta$ is an arbitrary phase, or in other words a connection
between the sine and cosine description. This however does not
mean that each and every trigonometric function is capable of
producing a viable description. Indeed, the case of a tangent
Gauss-Bonnet coupling or in general a hyperbolic trigonometric
case results in complex values for the scalar field. The most
promising choice would be the hyperbolic sine which produced a
real scalar field, however the model, dependent on $c$ and
$\gamma_1$, cannot produce compatible values for both the scalar
spectral index of primordial curvature perturbations and the
tensor-to-scalar ratio. In general, the same argument could be
used for the exponential choice or an exponential like choice for
the Gauss-Bonnet coupling. The choice of
$\xi(\phi)=\lambda_1(1-e^{\gamma_1\kappa\phi})$ is described only
by a complex scalar field and furthermore, the choice of a pure
exponential, meaning $\xi(\phi)=\lambda_1e^{\gamma_1\kappa\phi}$
results in eternal or no inflation since slow-roll index
$\epsilon_1$ turns out to be $\phi$-independent. In fact,
$\epsilon_1=\frac{1}{2}\left(\frac{4\kappa^2\gamma_1}{4\gamma_1^2\kappa^2-c}\right)^2$
which according to the choice of $\gamma_1$ and $c$ leads to
either eternal inflation for $\epsilon_1<1$ or no inflation for
$\epsilon_1>1$. It seems that this description is suitable for
trigonometric functions.
In principle there are many other models that may lead to a viable
phenomenology, however we refrain from discussing other examples
for brevity. The procedure is more or less the same, and nothing
new is added to our argument. In the next section, we shall
discuss another interesting case, the case that the scalar field
evolves dynamically in a constant-roll way.


\section{Inflationary Phenomenology of the GW170817-compatible Non-minimal Kinetic Coupling Corrected Einstein-Gauss-Bonnet Gravity: Constant-Roll Evolution}

In this procedure, we shall study a different scenario compared to
the slow-roll assumption for the scalar field. Particularly, we
shall assume that the scalar field evolves dynamically in a
constant-roll way
\cite{Inoue:2001zt,Tsamis:2003px,Kinney:2005vj,Tzirakis:2007bf,
Namjoo:2012aa,Martin:2012pe,Motohashi:2014ppa,Cai:2016ngx,
Motohashi:2017aob,Hirano:2016gmv,Anguelova:2015dgt,Cook:2015hma,
Kumar:2015mfa,Odintsov:2017yud,Odintsov:2017qpp,Lin:2015fqa,Gao:2017uja,Nojiri:2017qvx,Oikonomou:2017bjx,Odintsov:2017hbk,Oikonomou:2017xik,Cicciarella:2017nls,Awad:2017ign,Anguelova:2017djf,Ito:2017bnn,Karam:2017rpw,Yi:2017mxs,Mohammadi:2018oku,Gao:2018tdb,Mohammadi:2018wfk,Morse:2018kda,Cruces:2018cvq,GalvezGhersi:2018haa,Boisseau:2018rgy,Gao:2019sbz,Lin:2019fcz,Odintsov:2019ahz},
and as we shall see most of the equations we used previously,
shall remain the same, hence for simplicity we shall present only
the  equations that differ from the previous case. Essentially,
the constant-roll assumption states that,
\begin{equation}\label{constantroll}
\ddot\phi=\beta H\dot\phi\, ,
\end{equation}
where $\beta$ denotes the constant-roll parameter. As a result,
the overall phenomenology experiences certain changes. Firstly,
the equation for the scalar field shall be rewritten. Recalling
Eq. (\ref{difeq}), in view of the constant-roll condition
(\ref{constantroll}) we have,
\begin{equation}
\centering \label{dotphi2}
\dot\phi=\frac{4H(1-\beta)\xi'}{4\xi''+c\xi}\, .
\end{equation}
Here, it is worth making some comments. Firstly, this formula is
exact as no approximations were made. Secondly, for $\beta\ll 1$,
one obtains the same equation as in Eq. (\ref{dotphi}). Finally,
as expected, for $c\to 0$, we obtain the same formula as in the
case of Einstein-Gauss-Bonnet theory, either for the slow-roll or
constant-roll assumption respectively. Concerning the equations of
motion, they remain exactly the same, from
(\ref{motion1})-(\ref{motion3}). Furthermore, we shall assume the
exact same approximations, meaning the slow-roll conditions and
also the ones made in Eqs. (\ref{slowroll}) and
(\ref{extraconditions}). Thus, the simplified equations of motion
read,
\begin{equation}
\centering
\label{motion7}
H^2=\frac{\kappa^2V}{3}\, ,
\end{equation}
\begin{equation}
\centering \label{motion8} \dot H=-\frac{\kappa^2
H^2}{2}\left(\frac{4(1-\beta)\xi'}{4\xi''+c\xi}\right)^2\, ,
\end{equation}
\begin{equation}
\centering \label{motion9}
V'+3(1-\beta)(1+\frac{\beta}{3})H^2\frac{4\xi'}{4\xi''+c\xi}=0\, .
\end{equation}
The slow-roll indices do not experience any dramatic changes,
except of course for the slow-roll index $\epsilon_2$. In fact,
the same forms as before are still applicable with the only
difference being a factor of $1-\beta$. The same applies to the
auxiliary parameters $Q_i$ however we shall only showcase the
slow-roll indices since are more important. In fact, concerning
the slow-roll indices, the only significant difference is found as
expected in the slow-roll index $\epsilon_2$, which in the case at
hand is $\epsilon_2=\beta$. The slow-roll indices under the
constant-roll assumption are shown below,
\begin{equation}
\centering \label{index1b}
\epsilon_1=\frac{1}{2}\left(\frac{4(1-\beta)\kappa\xi'}{4\xi''+c\xi}\right)^2\,
,
\end{equation}
\begin{equation}
\centering
\label{index2b}
\epsilon_2=\beta\, ,
\end{equation}
\begin{equation}
\centering
\label{index4b}
\epsilon_4=\frac{2(1-\beta)\xi'}{4\xi''+c\xi}\frac{E'}{E}\, ,
\end{equation}
\begin{equation}
\centering
\label{index5b}
\epsilon_5=\frac{-16(1-\beta)\frac{\xi'^2H^2}{4\xi''+c\xi}+4c\xi H^2\left(\frac{4(1-\beta)\xi'}{4\xi''+c\xi}\right)^2}{\frac{1}{\kappa^2}-32(1-\beta)\frac{\xi'^2H^2}{4\xi''+c\xi}+32c\xi H^2\left(\frac{(1-\beta)\xi'}{4\xi''+c\xi}\right)^2}\, ,
\end{equation}
\begin{equation}
\centering \label{index6b}
\epsilon_6=-\frac{4H^2\frac{4\xi'^2(1-\beta)}{4\xi''+c\xi}(\frac{4\xi''(1-\beta)}{4\xi''+c\xi}+\beta-\epsilon_1)+cH^2\left(\frac{4\xi'(1-\beta)}{4\xi''+c\xi}\right)^2\left(\frac{2\xi'^2(1-\beta)}{4\xi''+c\xi}+\xi\beta\right)}{\frac{1}{\kappa^2}-32\frac{\xi'^2(1-\beta)}{4\xi''+c\xi}H^2+c\xi H^2\left(\frac{4\xi'(1-\beta)}{4\xi''+c\xi}\right)^2}\, .
\end{equation}
In this case as well, it is clear that the first two slow-roll
indices are simple while the rest are perplexed. Also, only a
factor of $1-\beta$ is now emergent, in contrast to the slow-roll
case, as mentioned before. Lastly, the $e$-foldings number is also
slightly changed in the constant-roll case, as is shown below,
\begin{equation}
\centering \label{efolds2}
N=\frac{1}{1-\beta}\int_{\phi_i}^{\phi_f}{\frac{4\xi''+c\xi}{4\xi'}d\phi}\,
.
\end{equation}
It can easily be inferred that the overall phenomenology is
similar to the one acquired previously. In particular, for
$\beta\ll 1$, the results are quite similar. In the following we
shall study a specific model and examine whether the constant-roll
assumption in general can produce compatible results. We shall
present an interesting example with the Gauss-Bonnet scalar
coupling being chosen to be a linear function of the scalar field.

\subsection{The Choice Of A Linear Gauss-Bonnet Coupling}

Let us assume a simple form for the Gauss-Bonnet scalar coupling
function, and particularly, let $\xi(\phi)$ be equal to,
\begin{equation}
\centering
\label{xiC}
\xi(\phi)=\lambda_2\kappa\phi\, ,
\end{equation}
so $\xi(\phi)$ is a linear function of the scalar field. In this
case, the second derivative with respect to the scalar field is
zero, meaning $\xi''=0$. As we shall show shortly, the condition
$\xi''$ being equal to zero does not affect dramatically the
phenomenology, on the contrary it makes our study easier, since
$\dot\phi$ is simplified. The corresponding scalar potential also
has a power-law form, and it can easily be found for the linear
scalar coupling, and it reads,
\begin{equation}
\centering \label{VC}
V(\phi)=V_2(\kappa\phi)^{-4\frac{(1-\beta)\kappa^2 }{c}}\, .
\end{equation}
It is clear that the exponent is once again general and is not
necessarily integer. Let us now proceed with the slow-roll
indices, Due to the linear choice of the scalar coupling function,
we have,
\begin{equation}
\centering \label{index1C} \epsilon_1=2
\left(\frac{2(1-\beta)\kappa}{c\phi}\right)^2\, .
\end{equation}
Due to the simple form of the slow-roll index $\epsilon_1$, the
initial and final value of the scalar field are easily obtained
and in this case are written as,
\begin{equation}
\centering
\label{phiiC}
\phi_i=\pm\sqrt{\frac{c\phi_f^2+8N(1-\beta)^2}{-c}}\, ,
\end{equation}
\begin{equation}
\centering \label{phifC} \phi_f=\pm\frac{2\kappa(1-\beta)\sqrt{2
}}{c}\, .
\end{equation}
The resulting theory can be compatible with the observational data
for a wide range of values of the free parameters. For example by
choosing ($\lambda_2$, $V_2$, $N$, $c$, $\beta$)=($10^{-8}$, 10,
60, -5, 0.009) in reduced Planck units ($\kappa^2=1$),  then the
observational indices read $n_S=0.968604$, $n_T=-0.0066667$ and
$r=0.0531572$ which are obviously compatible with the data coming
from the Planck 2018 collaboration. Moreover, the scalar field
decreases with time as $\phi_i=9.72595$ whereas $\phi_f=0.560594$
in Planck units, the sound wave velocity is equal to unity as
expected and in addition, the slow-roll conditions
 seem to be valid since the numerical values of the slow-roll indices are of order
$\mathcal{O}(10^{-3})$ and lesser. In particular,
$\epsilon_1=0.0033223$, $\epsilon_4=1.2\cdot10^{-6}$,
$\epsilon_5=-6.4\cdot10^{-8}$ and $\epsilon_6=5\cdot10^{-11}$.
\begin{figure}[h!]
\centering
\label{plot3}
\includegraphics[width=17pc]{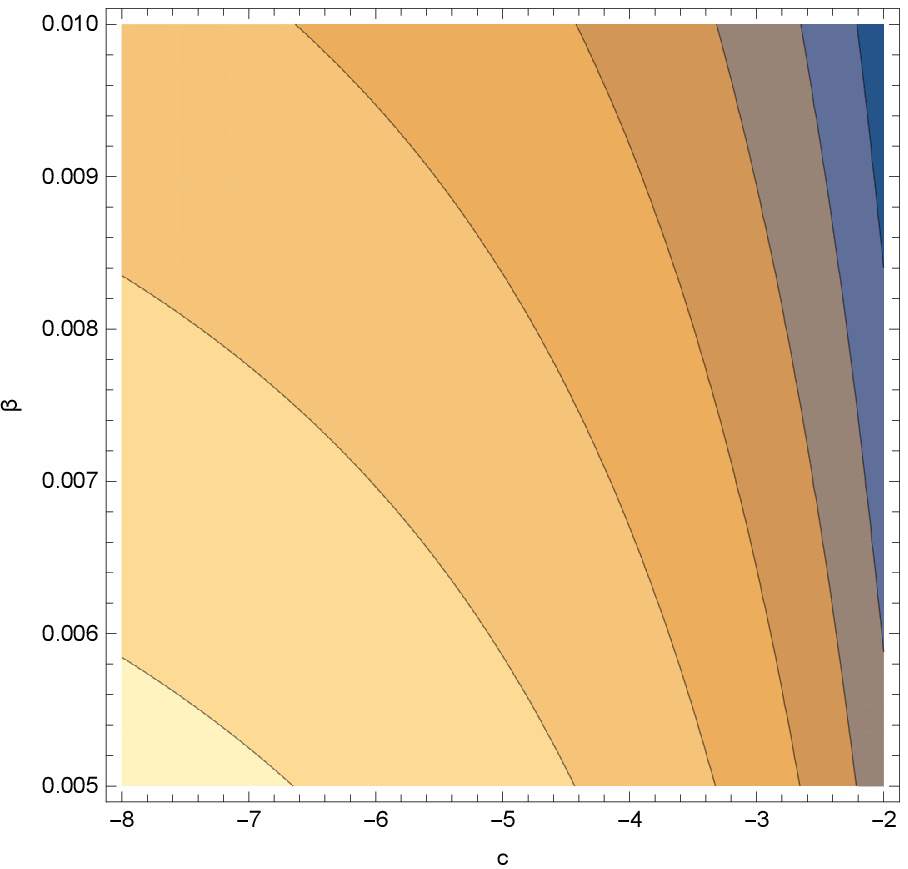}
\includegraphics[width=3pc]{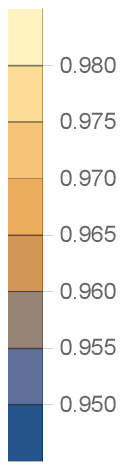}
\includegraphics[width=17pc]{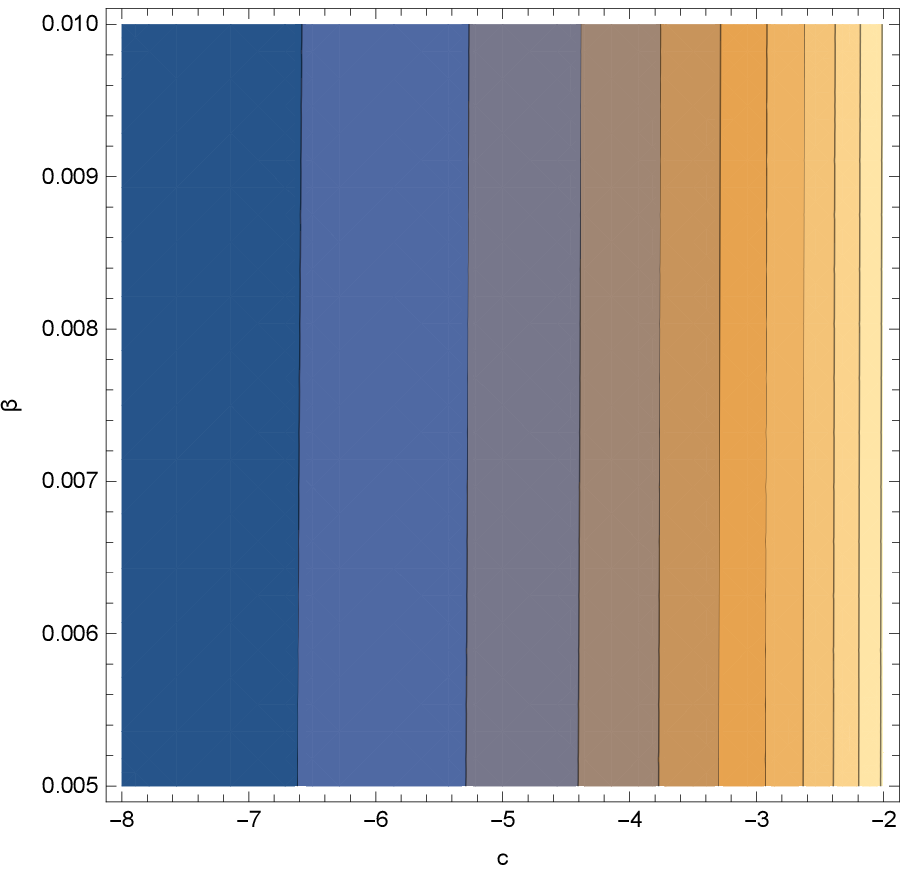}
\includegraphics[width=2.1pc]{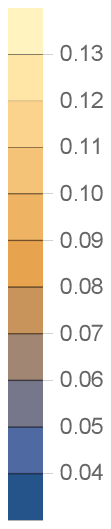}
\caption{The scalar spectral index $n_S$ (left) and
tensor-to-scalar ratio $r$ (right) with respect to constant-roll
parameter $\beta$ and the parameter $c$. Their values range from
[0.005, 0.01] and [-8,-2] respectively. It becomes apparent that
spectral index $n_S$ depends on both the free parameters, whereas
$r$ is only affected by $c$. In particular, the observational
indices are very sensitive since in such small area of designation
for the free parameters, the possible values for $n_s$ and $r$
have a wide range.}
\end{figure}
In this case, the scalar potential, which follows a power-law form
as well, has an exponent which is approximately equal to $0.793$,
hence once again it is not integer as expected. Perhaps there
exists a complete different choice for the free parameters which
produce a more familiar exponent. In Fig. \ref{plot3} we present
the dependence of the observational indices on the constant-roll
parameter $\beta$ and the parameter $c$. In general, these are not
the only parameters which affect the results. Indeed, $\lambda_2$
also affects the overall phenomenology as $\lambda_2=0.0001$ for
instance produces incompatible results, namely $n_S=0.944381$ and
$r=0.0751259$. The coefficient of the scalar potential also
affects the results. Here, increasing $V_2$ affects mildly the
scalar spectral index of the primordial curvature perturbations.
On the other hand, the tensor-to-scalar ratio varies since for
$V_2=10^3$, we have $r=0.0533064$ but for $V_2=10^5$, while $n_S$
is altered little, the tensor-to-scalar ratio is increased by an
order, in particular $r=0.115128$, hence it is necessary to choose
this value wisely. Referring to Fig. \ref{plot3} and in particular parameter $\beta$,
 it becomes apparent that there exist multiple values and especially of order
 $\mathcal{O}(10^{-2})$ and greater which are capable of producing  compatible with
the observations results however in our initial example we chose
to work with a value of $\mathcal{O}(10^{-3})$ for the
constant-roll parameter so as to obtain a slow-roll index
$\epsilon_2$ of equal order as the first at least. Decreasing the
order is obviously viable however it gets closer to the slow-roll
case hence the selection of $\beta=0.009$. Now the question is, is
there any dividing line between slow-roll and constant-roll
regimes? The constant-roll condition in general generates
non-Gaussianities, so in principle if a constant-roll regime
actually occurred, this could have observable effects, that may be
observed in the future. To our opinion, a constant-roll phase
occurs when $\beta$ is not a small parameter ($\beta\ll 1$), so
possibly when $\beta \mathcal(O)(10^{-3})$ or larger a
constant-roll regime is realized, this is why we chose $\beta$
taking the value $\beta=0.009$, as we discussed previously. For
smaller values of $\beta$, the slow-roll regime would be
approached.

Since an analysis referring to the sound wave velocity was made in the
slow-roll case, it is only fair to examine the impact of the combination
$c\xi$ in the constant-roll case as well. As it was shown before, this
particular choice of parameters results in $c_A=1$. Suppose now that
$\lambda_2\sim\mathcal{O}(10^{-3})$ and $c\sim\mathcal{O}(10^2)$.
As a result, the sound wave velocity during the ending stage of the inflationary
era obtains a numerical value of $c_A=1.0996+0.0044i$, which is complex,
therefore ghost instabilities are unavoidable. When $\lambda_2$ is decreased
however, no instabilities are present in our description. Moreover, such an
increase in parameters leads in a scalar spectral index of $n_S=-4.53$ which
is once again an insane value. This proves once and for all that the choice of
$c$ and $\lambda_2$ cannot be left unchecked. In principle, there could exist a
pair of values for the free parameters such that $c$ and $\lambda_2$ are close
to unity and compatibility is achieved, however to our knowledge, no such pair
was found. And for consistency, even if such pair was to be found, there is
a great chance that the model is infested with ghost instabilities or superluminal
velocities. In conclusion, the safest way to obtain a viable phenomenology
is to control the values of $c$ and $\lambda_2$ and constraining them to be small
many orders lesser than unity.

Finally, let us discuss here and validate whether the
approximations assumed in the section II hold true. Firstly, by
choosing ($\lambda_2$, $V_2$, $N$, $c$, $\beta$)=($10^{-8}$, 10,
60, -5, 0.009), we have $\dot H\sim\mathcal{O}(10^{-2})$ compared
to $H^2\sim\mathcal{O}(10)$, so the condition $\dot{H}\ll H^2$
holds true. Similarly, $\frac{1}{2}
\dot\phi^2\sim\mathcal{O}(10^{-2})$ while $V\sim\mathcal{O}(10)$,
so the assumption $\frac{1}{2} \dot\phi^2\ll V$ also holds true.
In addition, we have $24\dot\xi H^3\sim\mathcal{O}(10^{-5})$ and
$9c\xi\dot\phi^2H^2\sim\mathcal{O}(10^{-5})$ which are both
inferior to $V\sim\mathcal{O}(10)$, so the conditions in Eq.
(\ref{extraconditions}), hold true. Also, $16\dot\xi H\dot
H\sim\mathcal{O}(10^{-8})$,
$8(\ddot\xi-H\dot\xi)\sim\mathcal{O}(10^{-7})$ and
$c\left(2\xi\dot\phi^2\dot
H-6\xi\dot\phi^2H^2+4H\xi\dot\phi\ddot\phi+2H\dot\xi\dot\phi^2\right)\sim\mathcal{O}(10^{-5})$
while $ \dot\phi^2\sim\mathcal{O}(10^{-1})$. Lastly, we note that
$\xi'\mathcal{G}\sim\mathcal{O}(10^{-4})$,
$3c\left(H^2(\dot\xi\dot H+2\xi\ddot\phi)+2H(2\dot
H+3H^2)\xi\dot\phi\right)\sim\mathcal{O}(10^{-3})$ while
$V'\sim\mathcal{O}(1)$. Thus, all the approximations made are
indeed valid.

As a result, the kinetic coupling-corrected Einstein-Gauss-Bonnet
gravity with a linear coupling function, can lead to a viable
phenomenology assuming that graviton is massless. This is in
contrast to the simple GW170817-compatible Einstein-Gauss-Bonnet
gravity, which as we show in Ref. \cite{EPL}, a linear coupling
leads to the constant-roll condition inevitably, and the resulting
theory is phenomenologically questionable, with the regard to the
inflationary phenomenology. Actually, as we show in \cite{EPL},
the resulting scalar spectral index is always quite close to the
value $n_S\sim -1$.

\section{Conclusions}

In this work we studied non-minimal kinetic coupling corrected
Einstein-Gauss-Bonnet theories in view of the GW17017 event, which
restricted the gravitational wave speed to be equal to that of
light's. The constraint $c_T^2=1$ as we showed, constrained the
functional form of the scalar coupling and of the scalar
potential, and by using the slow-roll assumption, we showed how a
viable phenomenology can be obtained by the present theoretical
framework. We derived analytic formulas for the slow-roll indices
and for the observational characterizing the inflationary era, and
by using several illustrative examples, we demonstrated how the
compatibility of the inflationary phenomenology corresponding to
the present theory can be achieved. In principle several choices
of the scalar coupling function can also yield viable results, but
we refrained from presenting more models than a small class of
models, for brevity.

Our motivation of imposing the constraint $c_T^2=1$ in the
primordial gravitational waves, is mainly the fact that there is
no particle physics process related to the inflationary and
post-inflationary era that alters the graviton mass. The graviton
has a specific mass which can be determined by the string theory
governing the physics of the pre-inflationary epoch, so if the
graviton mass is specific during the inflationary era, there is no
fundamental reason that may allow it to change, at least to our
knowledge. As we showed, the constraint $c_T^2=1$ actually leads
to a phenomenologically viable non-minimal kinetic coupling
corrected Einstein-Gauss-Bonnet inflationary theory. Also, it is
worth mentioning that the constraint $c_T^2=1$ for the non-minimal
kinetic coupling corrected Einstein-Gauss-Bonnet also covers
theories with string correction terms of the form, results to the
same differential equation $\sim \xi(\phi)\Box\phi
g^{\mu\nu}\partial_\mu\phi\partial_\nu\phi$ and $\sim
\xi(\phi)\left(g^{\mu\nu}\partial_\mu\phi\partial_\nu\phi\right)^2$,
in which case the full effective Lagrangian is of the form,
\begin{equation}
\centering \label{ante_geia_protathlima_ston_peiraia}
\mathcal{S}=\int{d^4x\sqrt{-g}\left(\frac{R}{2\kappa^2}-\frac{1}{2}g^{\mu\nu}\partial_\mu\phi\partial_\nu\phi-V(\phi)-\xi(\phi)\mathcal{G}-c\xi(\phi)
G^{\mu\nu}\partial_\mu\phi\partial_\nu\phi-c_1\Box\phi
g^{\mu\nu}\partial_\mu\phi\partial_\nu\phi-c_2\left(g^{\mu\nu}\partial_\mu\phi\partial_\nu\phi\right)^2\right)}\,
.
\end{equation}
Theories of the form (\ref{ante_geia_protathlima_ston_peiraia})
belong to a wider class of Horndeski theories, so in this paper we
demonstrated how these theories may be revived formally, in view
of the GW170817 event. We did not study in detail however theories
of the form (\ref{ante_geia_protathlima_ston_peiraia}) since with
this paper we just wanted to demonstrate how the simplest class of
these theories can be revived. In a future work we might address
in detail the analysis of the full theory.

Finally, we discussed also the constant-roll scenario in the
context of the present theory. It is interesting to note that the
constant-roll condition is related to non-Gaussianities
\cite{Odintsov:2019ahz,Martin:2012pe}, so in a future work it
would be interesting to address this issue in detail in the
context of the non-minimal kinetic coupling corrected
Einstein-Gauss-Bonnet theories.

Finally, it is interesting to comment on the effects of the
constant-roll phase on the reheating era. This issue is
non-trivial to simply answer in a straightforward manner, and
needs to be addressed in detail in a future work, since for $F(R)$
gravity, a constant-roll phase alters the reheating temperature,
as was shown in Ref. \cite{Oikonomou:2017bjx}.

\end{document}